\documentstyle[aas2pp4,natbib209]{article}

\begin{document}

\title{Collisions of solid ice in planetesimal formation}

\author{J. Deckers\altaffilmark{1} and J. Teiser}
\affil{Fakult\"at f\"ur Physik, Universit\"at Duisburg-Essen, D-47057 Duisburg, Germany}

\altaffiltext{1}{johannes.deckers@uni-due.de}

\begin{abstract}
We present collision experiments of centimetre projectiles on to decimetre targets, both made up of solid ice, at velocities of $15\,\mathrm{m\,s^{-1}}$ to $45\,\mathrm{m\,s^{-1}}$ at an average temperature of $\mathrm{T_{avg}}=255.8\pm0.7\,\mathrm{K}$. In these collisions the centimetre body gets disrupted and part of it sticks to the target. This behaviour can be observed up to an upper threshold, that depends on the projectile size, beyond which there is no mass transfer. In collisions of small particles, as produced by the disruption of the centimetre projectiles, we also find mass transfer to the target. In this way the larger body can gain mass, although the efficiency of the initial mass transfer is rather low. These collision results can be applied to planetesimal formation near the snowline, where evaporation and condensation is expected to produce solid ice. In free fall collisions at velocities up to about $7\,\mathrm{m\,s^{-1}}$, we investigated the threshold to fragmentation and coefficient of restitution of centimetre ice spheres.
\end{abstract}

\section{Introduction}

Planets are formed by accretion of kilometre-sized objects, the planetesimals. Although observations as well as experiments and simulations have provided considerable insight into planetesimal formation, the processes involved are not yet entirely understood.\\
The models for planetesimal formation can roughly be divided into two main groups. One group considers growth through sticking in mutual collisions. Once aggregates reach the millimetre-to-centimetre size range, mutual collision do not necessarily lead to growth by sticking. These collisions can also result in bouncing \citep{zsom2010,jankowski2012,kelling2014} or even fragmentation \citep{beitz2011,deckers2013} of agglomerates, which stalls further growth.\\
However, in collisions of bodies of different sizes the larger body can gain mass, as part of the smaller body sticks to it after the collision \citep{kothe2010,teiser2011b,meisner2013,deckers2014}. This mass transfer is possible even at higher collision velocities.\\
Another group of models describes planetesimal growth by gravitational collapse in dense regions of the protoplanetary disc. Such high particle concentration could be achieved by turbulence \citep{johansen2006}, baroclinic instability \citep{lyra2011} or streaming instability \citep{chiang2010,johansen2007}.\\
Icy bodies and their collision dynamics are crucial for planetesimal formation \citep{okuzumi2012,drazkowska2014}, as water ice is a highly abundant material in the outer regions of protoplanetary discs. So far, a lot of the experimental work has focused on investigating collisions of dust and only a few studies analysed collisions of decimetre ice \citep{shimaki2012a,yasui2014}.\\
Ice is considered to be porous in most studies, simulations as well as experiments, that investigate the possibility of icy planetesimal formation \citep{shimaki2012a,kataoka2013}. Close to the snowline however, thermal processing of icy bodies will lead to compaction. \citet{ros2013} model the growth of icy bodies through condensation of water vapour on to already existing ice particles. In this way solid ice particles can grow up to sizes of decimetres, but their model does not consider sticking and fragmentation.\\
Therefore, we conducted collision experiments with centimetre projectiles impacting decimetre targets, all made up of solid water ice, at collision velocities from about $1\,\mathrm{m\,s^{-1}}$ to $45\,\mathrm{m\,s^{-1}}$.\\

\section{Experiment}

Collision experiments of solid centimetre ice projectiles impacting decimetre ice targets were conducted in a cold chamber at an average temperature of $\mathrm{T_{avg}}=255.8 \pm 0.7\,\mathrm{K}$.

\subsection{Setup for the Collision Experiments}

\begin{figure}[ht!]
\epsscale{1.}
\plotone{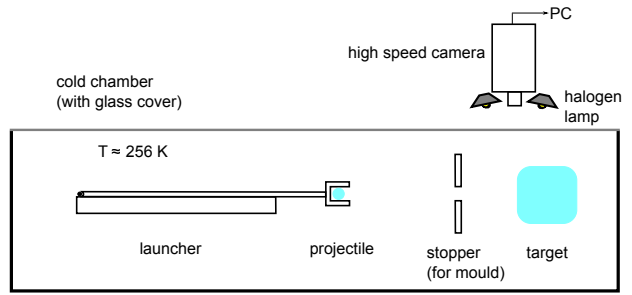}
\caption{Experimental setup for the collisions.\label{aufbau}}
\end{figure}

Fig. \ref{aufbau} shows the experimental setup for the collision experiments. The centimetre ice projectile is placed into the sample holder, an aluminium mould. It is then accelerated towards the target by a launcher.\\

The mould is stopped by a metal plate so that only the projectile impacts the target. A metal rail (not shown in Fig. \ref{aufbau} for simplicity) is attached to the stopper plate, so that the mould hits the opening correctly and the projectile can get through. Experiments in which the projectile gets disrupted before reaching the target are excluded from the analysis. In this way we conducted collisions of centimetre projectiles at velocities between about $15\,\mathrm{m\,s^{-1}}$ and $45\,\mathrm{m\,s^{-1}}$ and collisions of millimetre sized projectile fragments at velocities up to $50\,\mathrm{m\,s^{-1}}$.The projectiles hit the flat top of the cylindrical target (see Fig. \ref{bilder}). The collisions are then observed by a high speed camera at 5000 fps and illuminated by halogen lamps. Particles down to about $150\,\mathrm{\mu m}$ can be resolved by the camera.\\
In addition to that we conducted free fall collisions at velocities between $0.9\,\mathrm{m\,s^{-1}}$ and $6.5\,\mathrm{m\,s^{-1}}$ and investigated the fragmentation thresholds of the solid cm ice.\\

\subsection{Sample Preparation}

The samples were prepared by pouring demineralized water into a silicone form, which is then put into the cold chamber. The projectiles are spherical and have diameters of $1.7~\mathrm{cm}$, $2~\mathrm{cm}$, $2.5~\mathrm{cm}$ and $3.3~\mathrm{cm}$ (with mean masses of $2.28\pm0.07~\mathrm{g}$, $3.83\pm0.11~\mathrm{g}$, $7.48\pm0.16~\mathrm{g}$ and $16.83\pm0.63~\mathrm{g}$, respectively). The targets are cylinders with a diameter of $12~\mathrm{cm}$, a mean height of $12.9\pm1~\mathrm{cm}$ and mean mass of $1293\pm107\,\mathrm{g}$.\\
Both projectiles and targets look clear, with tiny air bubbles enclosed in them. Due to the air bubbles the projectiles have a small porosity of less than 5\%, assuming a density of ice of $\rho=0.92\,\mathrm{g\,cm^{-3}}$. This small porosity can also be found in the projectiles and targets used in the experiments of \citet{yasui2014}. The small porosity does not influence the outcome of their collision experiments, which we thus consider to be the case in our experiments as well.\\

\section{Results}
In order to apply the results of the experiments to planetesimal formation, the analysed threshold conditions have to be scaled to the lower ambient temperatures of protoplanetary discs (see discussion in Section \ref{plfo} for details).

\subsection{Threshold between collisions with and without mass gain}

\begin{figure}[ht!]
\epsscale{1.}
\plotone{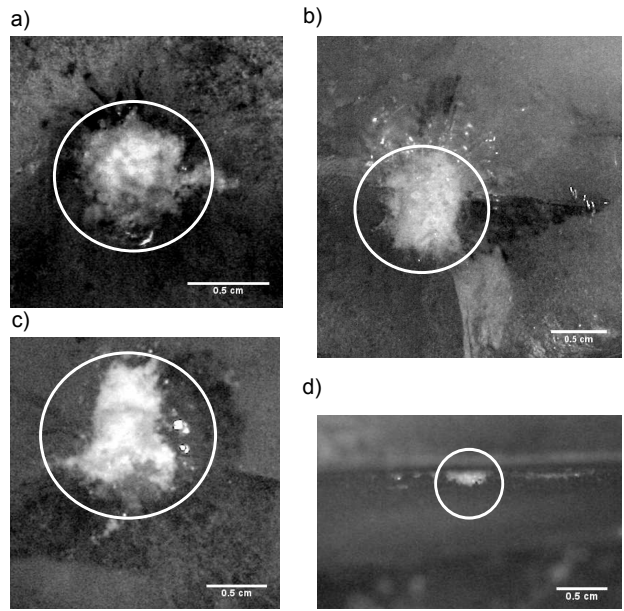}
\caption{View of targets after a collision. The white circles mark the grown structures on the target. Pictures in a) to c) show the top of the target, picture in d) shows a side view.\label{bilder}}
\end{figure}

\begin{figure}[ht!]
\epsscale{1}
\plotone{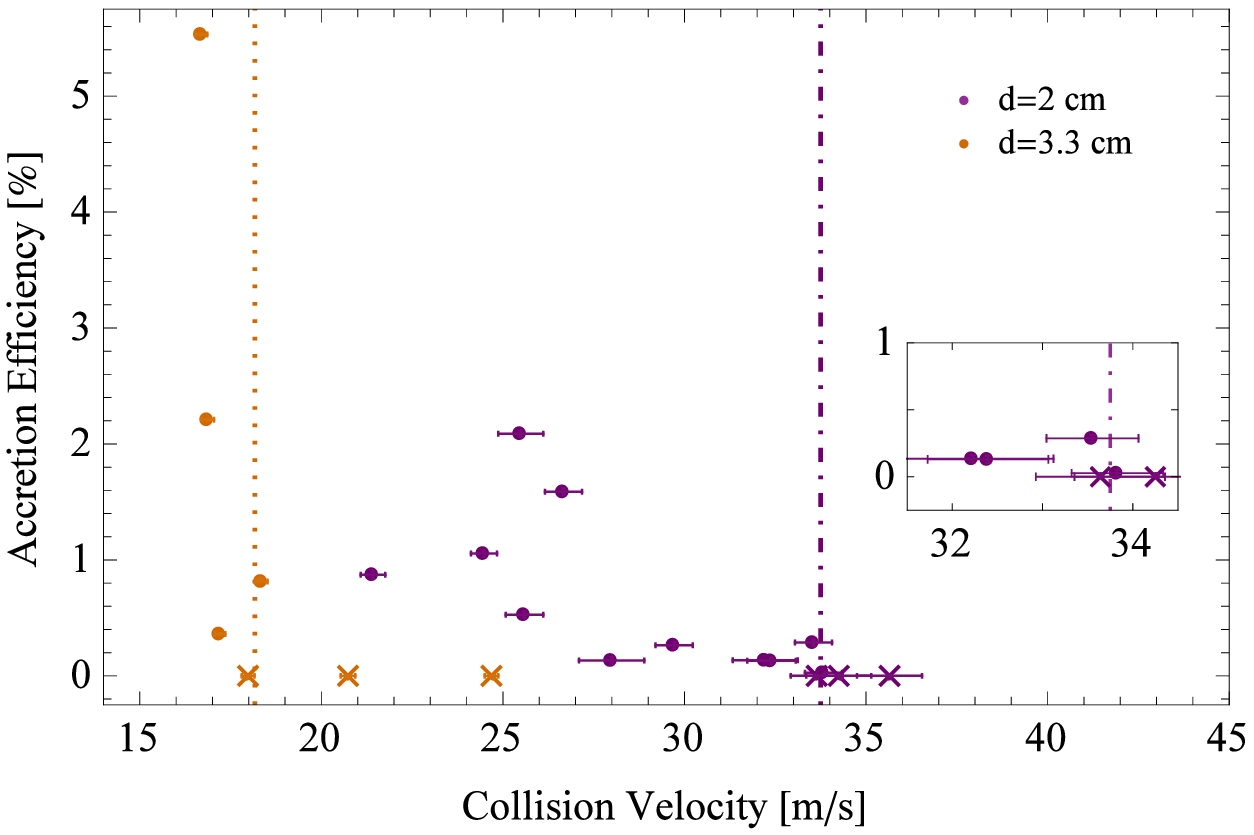}
\plotone{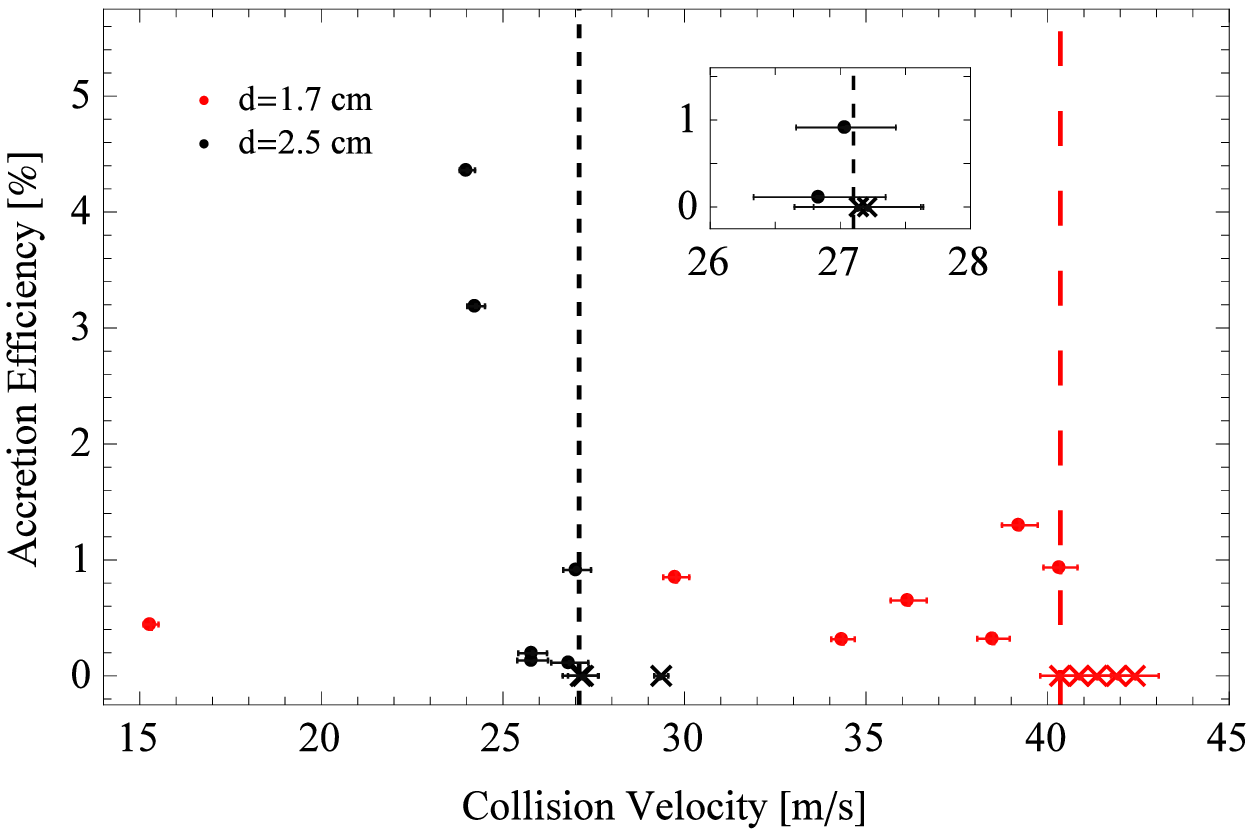}
\caption{Accretion efficiency for collisions of projectiles of different sizes. Circles show collisions with mass transfer, crosses collisions without mass transfer. Dashed lines show the thresholds. Collisions of spheres with $d=2~\mathrm{cm}$ and $d=3.3~\mathrm{cm}$ are shown in the top panel, those with $d=1.7~\mathrm{cm}$ and $d=2.5~\mathrm{cm}$ in the bottom panel.\label{grenze}}
\end{figure}

The collision velocity is calculated by tracking an edge of the projectile in the 2D camera images. In the collision experiments at higher velocities (experimental setup, see Fig. \ref{aufbau}), we observe two different outcomes of collisions. At lower velocities we observe mass transfer to the target, a small part of the projectile sticks firmly to the target after the collision. At higher velocities, there is no mass transfer.\\
The mass of the grown structure on the target can not be measured directly, as the structure sticks to the target firmly and can not be removed without destroying it. It can not be measured directly either, because the precision of a scale that is suited for the target mass of about $1300\,\mathrm{g}$ is not high enough to measure a mass difference down to a few $\mathrm{mg}$. We can thus only estimate the mass of the grown structure by taking pictures of it after the collision. Fig. \ref{bilder} shows examples of these pictures. Here, the contrast between the grown structure and the target is enhanced, so that the grown structure, marked by the white circles, is better visible. From these pictures the volume of the grown structure is calculated by taking the top and the side view. The accreted mass is calculated from the volume assuming that the density of the grown structure has not changed in the collision. In this way we get an accretion efficiency, calculated by normalizing the accreted mass to the mass of the projectile before the collision, of less than 6\% (see Fig. \ref{grenze}).\\

\begin{figure}[htb!]
\epsscale{1.}
\plotone{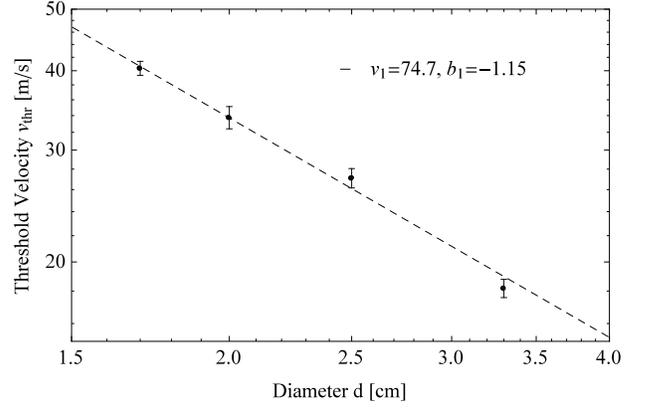}
\caption{Threshold velocities between collisions with and without mass transfer for projectiles of different diameter. The dashed line shows a fit using Equation \ref{power1}.\label{grenzv}}
\end{figure}

Fig. \ref{grenze} shows the accretion efficiency in the collisions of the different projectiles and the thresholds to collisions without mass gain, which are marked by the crosses. The threshold between the two collision outcomes (shown by the dashed lines in Fig. \ref{grenze}) depends on the projectile size. Fig. \ref{grenzv} shows an overview of the threshold velocities $v_{thr}$ for projectiles of different sizes. The dashed line shows a power law fit to the data using the equation

\begin{equation}
v_{thr}=v_1\cdot\left(\frac{d}{\mathrm{cm}}\right)^{b_1}\,\mathrm{m\,s^{-1}}.\label{power1}
\end{equation}

In both collision outcomes, the projectiles get disrupted in the collision, with a fragment distribution that follows a power law in the form of $N\propto r^b$, where $r$ is the radius of the particles and $N$ the number of particles in a finite sized bin centred at a radius $r_i$. This is similar to the distributions observed in collisions of centimetre on to decimetre dust agglomerates \citep{deckers2014}. We exemplarily analysed the fragment distribution of three collisions and found a mean exponent of $b=-3.6\pm0.2$.\\

\subsection{Dependence of the threshold on the impact angle}

\begin{figure}[htb!]
\epsscale{1.}
\plotone{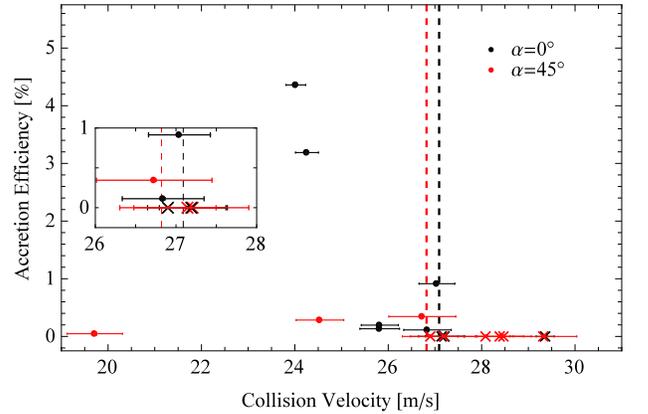}
\caption{Accretion efficiency for collisions of projectiles of $d=2.5\,\mathrm{cm}$ at an impact angle $\alpha$ of $0^{\circ}$ and $45^{\circ}$. Circles show collisions with mass transfer, crosses collisions without mass transfer. Dashed lines show the thresholds.\label{winkel}}
\end{figure}

To analyse whether the impact angle influences the threshold between collisions with and without mass gain, we conducted collision experiments with an impact angle of $45^{\circ}$. Fig. \ref{winkel} shows the comparison of the collision results of projectiles of $d=2.5\,\mathrm{cm}$ at impact angles of $0^{\circ}$ (taken from Fig. \ref{grenze}) and $45^{\circ}$.\\
For $\alpha=45^{\circ}$ the threshold lies at $v_{thr}=26.82\pm0.78\,\mathrm{m\,s^{-1}}$, which is within the margin of error identical to the threshold at $\alpha=0^{\circ}$ of $v_{thr}=27.09\pm0.51\,\mathrm{m\,s^{-1}}$. This suggests that at least for small angles the impact angle does not influence the threshold conditions significantly.\\

\subsection{Impacts of irregular ice particles}

\begin{figure}[htb!]
\epsscale{1.}
\plotone{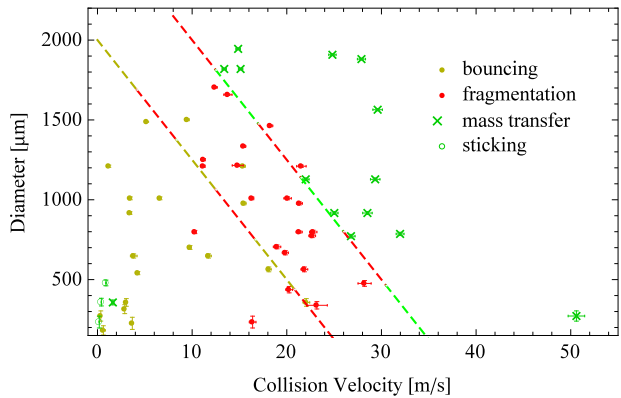}
\caption{Results of collisions of small particles of different sizes on to decimetre targets. The dashed lines illustrate the transition between different collision results.\label{trans}}
\end{figure}

We analysed the impacts of small particles of less than $2\,\mathrm{mm}$ in diameter on to decimetre targets to investigate the impacts of projectile fragments produced by the disruption of a centimetre projectile in a collision. The projectile diameter was calculated by determining the projected area from the camera images, assuming spherical particles. Fig. \ref{trans} shows the outcome of collisions of these small particles with decimetre targets at a range of velocities of $0.2\,\mathrm{m\,s^{-1}}$ to $50\,\mathrm{m\,s^{-1}}$.\\
At lower velocities most of the projectiles bounce off the target, with a few sticking collisions of small projectiles at velocities of less than $1\,\mathrm{m\,s^{-1}}$. With increasing velocity, beyond the yellow--red dashed line, fragmentation of the projectile becomes dominant. For even higher velocities, beyond the red--green line, we find mass transfer. In these collisions, a small part of the projectile sticks to the target after the collision, just like in the collisions of the centimetre projectiles on to the decimetre targets described earlier.\\

Fig. \ref{trans} also shows that the transition velocities between different collision outcomes decreases with increasing projectile size.\\

\subsection{Multiple impacts on to the same target}

\begin{figure}[htb!]
\epsscale{1.}
\plotone{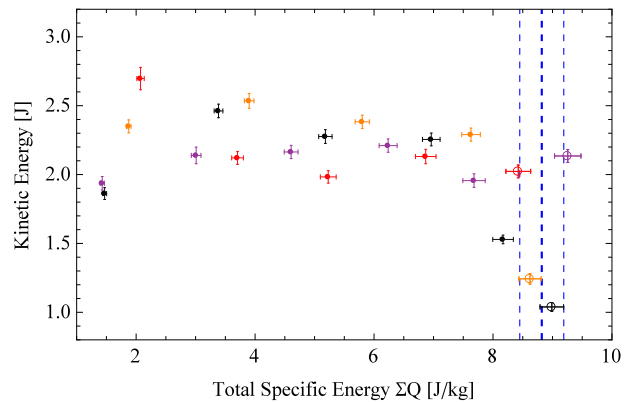}
\caption{Kinetic energy of the individual collisions plotted against the total specific kinetic energy. Symbols in the same colour are from collisions with the same target. The open circles mark the collisions, where erosion begins.\label{erosion}}
\end{figure}

We investigated the outcome of multiple impacts on to the same area of a target. The kinetic energy $E_{kin}$ of a collision is calculated from the collision velocity as $E_{kin}=1/2\,mv^2$. The error is calculated assuming an error in the projectile mass of $\Delta m=0.1\,\mathrm{g}$. The specific energy $Q$ is calculated by normalizing the kinetic energy of a collision by the mass of the collision partners.\\
Fig. \ref{erosion} shows the kinetic energy of the individual collisions plotted against the total specific energy $\Sigma Q$, which is calculated as the sum of the specific energy of a collision and all previous collisions. The open circles show where erosion of the multiply impacted spot of the target starts. In these collisions we observe only very minor damage to the target surface, with less than $0.2\%$ of the target mass getting eroded. The thick dashed line shows the mean specific energy needed for erosion of the target, $\Sigma Q=8.82\pm0.37\,\mathrm{J\,kg^{-1}}$, the thin dashed lines show the standard deviation. Table \ref{uebersicht_erosion} shows an overview of the results of the multiple impacts.\\

\begin{table}[htb!]
\centering
\begin{tabular}{|c||c|c|c|c|} \hline

Series nr.&1&2&3&4\\ \hline
Nr. of collisions&5&6&6&5\\ \hline
$\Sigma Q\,[\mathrm{J\,kg^{-1}}]$&8.43&9.26&8.99&8.62\\ \hline
\end{tabular}
\caption{Overview of the total specific energy $\Sigma Q$ needed for the erosion of a target.}
\label{uebersicht_erosion}
\end{table}

The total specific energy required for erosion is slightly lower, when only five collisions on to the same target are performed, than in the experiments with six collisions. The mean $\Sigma Q$ for five collisions is $8.52\,\mathrm{J\,kg^{-1}}$, for six collisions it is $9.12\,\mathrm{J\,kg^{-1}}$. This is in qualitative agreement with the findings of \citet{yasui2014}, who performed multiple impacts of centimetre projectiles on to decimetre targets, both made up of solid ice, at velocities of $84\,\mathrm{m\,s^{-1}}$ to $502\,\mathrm{m\,s^{-1}}$. They show that the total specific energy needed to catastrophically disrupt a target increases with the number of collisions, from $74\,\mathrm{J\,kg^{-1}}$ in a single collision to $112\,\mathrm{J\,kg^{-1}}$ in four collisions (see table 2 in \citealt{yasui2014}).\\ 

\subsection{Fragmentation threshold and coefficient of restitution of cm ice}

We analysed collisions of projectiles with $d=2~\mathrm{cm}$ and $d=2.5~\mathrm{cm}$ on to decimetre targets at lower velocities (from $0.9\,\mathrm{m\,s^{-1}}$ to $6.5\,\mathrm{m\,s^{-1}}$). These collisions were conducted by placing the projectile on to a pre-cooled ejection mechanism, the one that was already used in previous experiments \citep{deckers2014}, and then letting it drop from various heights.\\

\begin{figure}[htb!]
\epsscale{1.}
\plotone{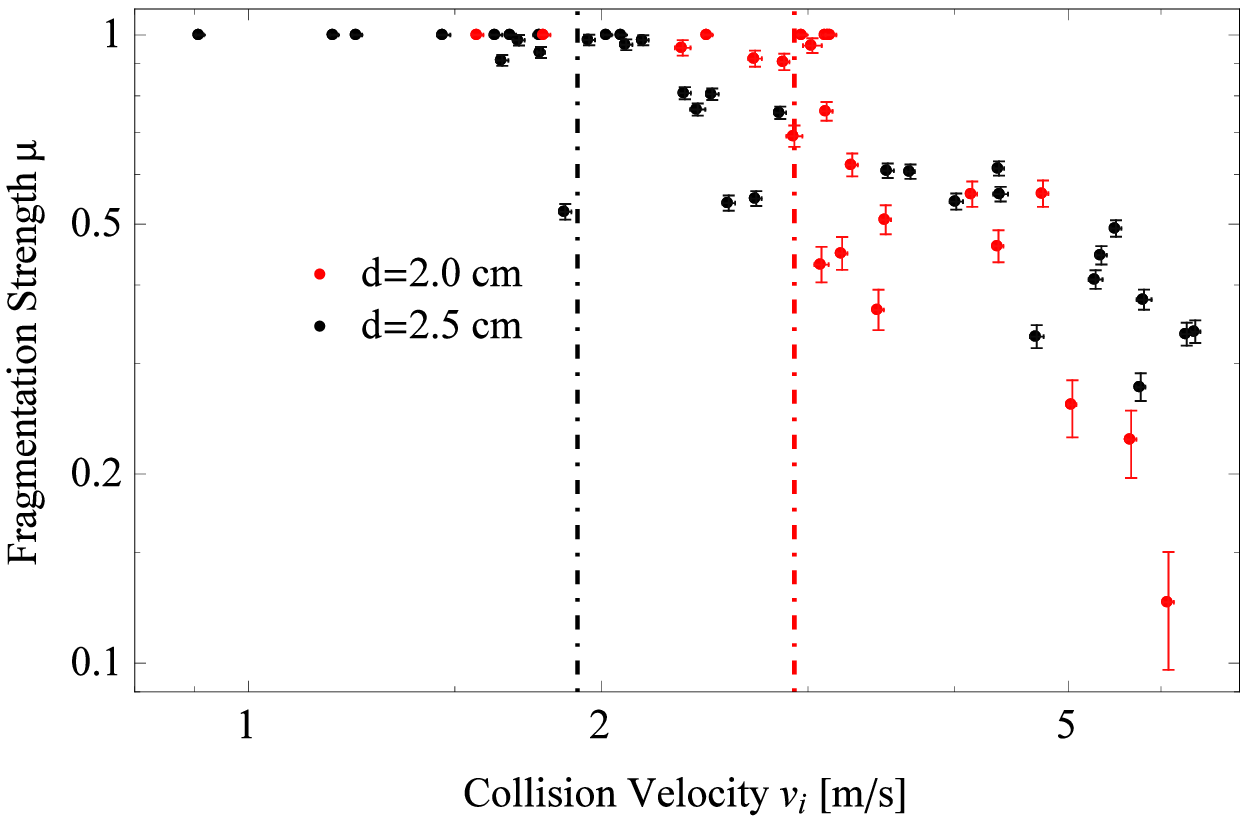}
\plotone{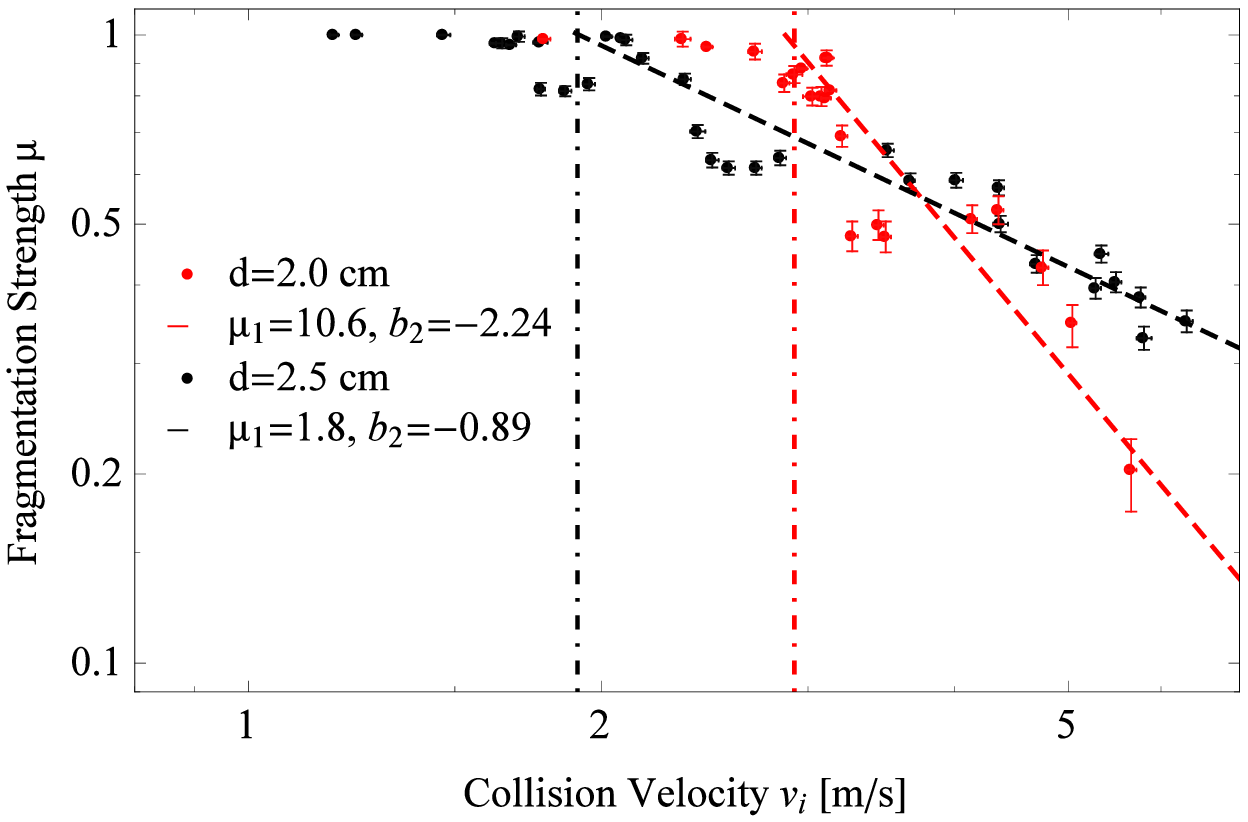}
\caption{Fragmentation strength of ice projectiles with $d=2~\mathrm{cm}$ and $d=2.5~\mathrm{cm}$. The top plot shows the collision results, the bottom plot the averaged mean over three data points. The dotdashed lines show the threshold to fragmentation, the dashed lines fits using Equation \ref{power2}.\label{klein}}
\end{figure}

Fig. \ref{klein} shows the fragmentation strength $\mu=M_f/M_0$, where $M_f$ is the mass of the largest projectile fragment and $M_0$ its original mass, plotted against the collision velocity. The dotdashed lines show the threshold to fragmentation, which is here defined by $\mu < 0.95$ (as done by \citealt{higa1998} and \citealt{yasui2014}). The threshold velocity is calculated by taking the mean value between the highest velocity with $\mu > 0.95$ and the lowest velocity with $\mu < 0.95$. The threshold is at $2.92\pm0.34\,\mathrm{m\,s^{-1}}$ ($d=2~\mathrm{cm}$) and $1.91\pm0.39\,\mathrm{m\,s^{-1}}$ ($d=2.5~\mathrm{cm}$).\\
The bottom plot in Fig. \ref{klein} shows the moving average of the fragmentation strength, an average of three data points. The dashed lines show fits to the data with velocity $v_i$ larger than the threshold velocity. Here, we use a power law fit, as used by \citet{davis1990} and \citet{arakawa1999}, in the form of 

\begin{equation}
\mu=\mu_1\cdot \left(\frac{v_i}{\mathrm{m\,s^{-1}}}\right)^{b_2}.\label{power2}
\end{equation}

\begin{figure}[htb!]
\epsscale{1.}
\plotone{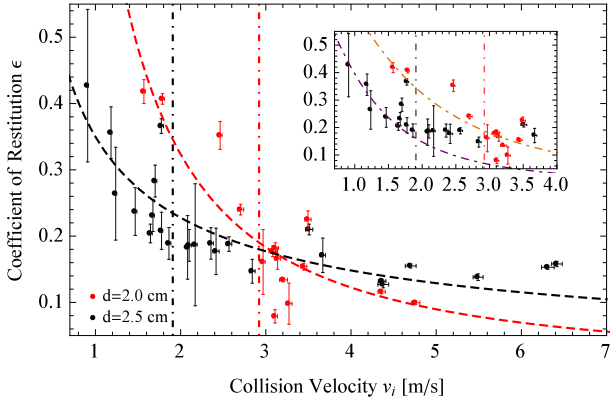}
\caption{Coefficient of restitution in collisions of ice projectiles with $d=2\,\mathrm{cm}$ and $d=2.5\,\mathrm{cm}$. The dashed lines show fits using Equation \ref{power3}, with $\epsilon_1=0.85$, $b_3=-1.4$ ($d=2\,\mathrm{cm}$) and $\epsilon_1=0.35$, $b_3=-0.62$ ($d=2.5\,\mathrm{cm}$). The inset shows fits using Equation \ref{restmodell}, with $\epsilon_{eq}=0.79$, $v_c=0.49\,\mathrm{cm\,s^{-1}}$ ($d=2\,\mathrm{cm}$) and $\epsilon_{eq}=0.68$, $v_c=0.36\,\mathrm{cm\,s^{-1}}$ ($d=2.5\,\mathrm{cm}$).\label{rest}}
\end{figure}

Fig. \ref{rest} shows the coefficient of restitution $\epsilon=v_f/v_i$, the ratio of the velocity after the collision $v_f$ and before the collision $v_i$ for most of the collisions shown in Fig. \ref{klein}. In some collisions it is not possible to determine the coefficient of restitution, e.g. because the projectile mainly rotates after the collision or rolls over the target surface. The angular velocity of the projectiles can not be calculated reliably from the 2D images, as there are no significant points on the smooth surface of the spheres to determine the rotation. Projectiles do not rotate before the collision and have no significant rotation in the collisions analysed here. For fragmenting collisions, $\epsilon$ was calculated by taking the velocity of the largest fragment after the collision. The vertical dot--dashed lines show the threshold to fragmentation, as in Fig. \ref{klein}.\\

The dotdashed lines in the inset of Fig. \ref{rest} show fits to the data using the equation

\begin{equation}
\epsilon=\epsilon_{eq}\cdot \mathrm{e}^{c(Ln(\frac{v_i-v_s}{v_c}))^2},\label{restmodell}
\end{equation}

where $\epsilon_{eq}$ is the maximum coefficient of restitution, $v_c$ the threshold to inelastic collisions, $v_s$ the sticking velocity and $c$ a fitting parameter. This equation comes from an analytical model developed by \citet{musiolik2015} and is similar to the empirical dependence found by \citet{higa1998}. As we do not observe sticking in our experiments and \citet{musiolik2015} show that $v_s \ll v_c$, we assume $v_s=0$. For the fits in Fig. \ref{rest} we get the following values for the parameters: $\epsilon_{eq}=0.79$, $v_c=0.49\,\mathrm{cm\,s^{-1}}$ and $c=-0.45$ ( for $d=2\,\mathrm{cm}$) and $\epsilon_{eq}=0.68$, $v_c=0.36\,\mathrm{cm\,s^{-1}}$ and $c=-0.52$ (for $d=2.5\,\mathrm{cm}$).\\
The model of \citet{musiolik2015} allows us to determine the maximum coefficient of restitution $\epsilon_{eq}$ and the threshold to inelastic collisions $v_c$, which are interesting for the collision dynamics of the ice spheres. The model fits well to the data up to the threshold to fragmentation, but fails to describe the relatively high constant values of $\epsilon$ observed for the larger spheres at higher velocities. Therefore, we fit the data over the whole range of velocities with a power law, shown by the dashed lines in Fig. \ref{rest}, following the equation

\begin{equation}
\epsilon=\epsilon_1\cdot \left(\frac{v_i}{\mathrm{m\,s^{-1}}}\right)^{b_3}.\label{power3}
\end{equation}

\section{Discussion}

\subsection{Collisions with mass transfer}

In the collision experiments of centimetre projectiles impacting decimetre targets we find mass transfer at lower collision velocities. A small part of the projectile, less than $6\%$ of its mass, sticks firmly to the target. At higher velocities there is no mass transfer. The threshold velocity decreases with increasing projectile size. This is in good agreement with the experiments of \citet{higa1998}, who found a decrease in the threshold velocity between elastic and inelastic collisions with increasing projectile size.\\
There is no obvious explanation for this upper threshold. Mass transfer in collisions of dust agglomerates by \citet{deckers2014} suggests that the part of the agglomerate that hits the target first will stick to it. The rest of the porous dust projectile then 'flows' around this structure forming it into a cone. This idea is supported by a comparison to the behaviour of a jet of granular matter, which forms a cone when impacting a target that is of comparable size to the jet \citep{cheng2007}. Solid ice projectiles apparently behave differently. At higher velocities the part of the projectile that sticks to the target initially might get disrupted by the material impacting the target after that. The fact that the transition velocity decreases with increasing projectile size gives an additional hint that this might be the case.\\
Fig. \ref{winkel} shows no dependence of the threshold on the impact angle for impact angles up to $45^{\circ}$. This indicates that the collision geometry might not be the important parameter that determines the outcome of a collision. This is in contrast to the experiments of \citet{deckers2014}, who find that in collisions of porous dust agglomerates, the accretion efficiency decreases significantly with increasing impact angle, even for smaller impact angles. Apart from the collision velocity, it is most probably the surface of both the projectile as well as the target that plays a crucial role. As we only made experiments with smooth and frost- free bodies, we can not make any analysis of the influence of the roughness or the coating of the surface of the collision partners. It is important to note that the impact angle will most probably have a significant influence on the collision outcome, as collisions become glancing. But this is beyond the scope of this study, as we can not perform collisions at higher impact angles with the present setup.\\

\subsection{Low velocity collisions}

In free fall collisions, we investigated the collisions of centimetre ice spheres at velocities of up to $6.5\,\mathrm{m\,s^{-1}}$ and analysed the threshold to fragmentation and the coefficient of restitution. Here, we used projectiles with diameters of $2\,\mathrm{cm}$ and $2.5\,\mathrm{cm}$.\\
The threshold to fragmentation is lower for the bigger spheres, which again is in good agreement with the experiments of \citet{higa1998}.\\
The coefficient of restitution decreases up to the threshold velocity to fragmentation and is more or less constant after that. \citet{higa1998} analysed the coefficient of restitution for collisions of solid ice projectiles of various sizes, with diameters ranging from $0.28\,\mathrm{cm}$ to $7.2\,\mathrm{cm}$, on to decimetre ice blocks. They empirically found a dependence of the coefficient of restitution onto the collision velocity. We use a numerical model developed by \citet{musiolik2015}, which yields a dependence that is similar to the one empirically found by \citet{higa1998}. \citet{musiolik2015} investigated the collisions of $\mathrm{CO_2}$ ice particles with an average size of about $180\,\mathrm{\mu m}$ at velocities up to $2\,\mathrm{m\,s^{-1}}$. Their model is based on the observation that the coefficient of restitution is symmetric, i.e. it increases from sticking collisions at very low velocities to elastic collisions at about $0.25\,\mathrm{m\,s^{-1}}$ and then decreases for increasing velocity. This modified dependence fits quite well to our data.\\
In contrast to these experiments, \citet{heisselmann2010} find no dependence of the coefficient of restitution on the impact velocity. \citet{heisselmann2010} investigated mutual collisions of solid ice spheres with diameters of $1.5\,\mathrm{cm}$ at velocities from $0.06\,\mathrm{m\,s^{-1}}$ to $0.22\,\mathrm{m\,s^{-1}}$. \citet{heisselmann2010} suggest that partial surface coverage by frost could be one reason for the spread in the coefficients of restitution observed in their experiments. In collisions of millimetre sized ice particles at velocities from $0.26\,\mathrm{m\,s^{-1}}$ to $0.51\,\mathrm{m\,s^{-1}}$, \citet{hill2015} do not find any correlation between the coefficient of restitution and the impact velocity, impact parameter or the ambient temperature, either. \citet{hill2015} state that the reason for the large spread in the measured coefficients of restitutions is the surface roughness of the particles.

\subsection{Astrophysical application - planetesimal formation\label{plfo}}

Several recent studies investigate the possibilities of icy planetesimal formation, e.g. \citet{okuzumi2012} or \citet{kataoka2013}. One idea behind these studies is that the increased stickiness of ice in comparison to dust might help to overcome bouncing and fragmentation that potentially stall the growth of dust agglomerates in the centimetre to decimetre regime. \citet{kataoka2013} analyse coagulation of highly porous ice, considering a porosity evolution in which agglomerates get compressed in collisions as well as by gas and self-gravitational compression, but remain highly porous up to a size of about $100\,\mathrm{m}$. Porous bodies can, however, become compact by thermal processing, such as sintering. Thermal processing has been widely discussed for dust, metals as well as silicates, but can also affect water ice. \citet{ros2013} propose growth by condensation of water vapour that diffuses over the snowline due to the turbulent motion of the gas, on to existing ice particles. In their model millimetre particles can grow to decimetre bodies in this way. These processes form solid icy bodies that might contain a dust nucleus. \citet{sirono2011} analyses sintering processes on ice-covered dust particles through sublimation and condensation. Inward drifting icy bodies start to sublimate as they approach the snowline. \citet{sirono2011} shows that molecules condense on to the necks connecting single grains, which significantly changes their collision properties. First attempts of the Mupus instrument on Philae, the lander of the Rosetta mission investigating comet 67P (Churyumov--Gerasimenko), to drill into the surface of the comet were not successful. This might indicate that the comet partly consists of solid ice.\\
In the minimum mass solar nebula (MMSN) model, the temperature close to the snowline, which is at about 3 AU from the sun in the MMSN model, is at about $160\,\mathrm{K}$ \citep{hayashi1985}. This is much lower than the ambient temperature of $256\,\mathrm{K}$ in the experiments presented here. In order to apply the results of our experiments to planetesimal formation it is thus important to know, how the reduced temperature influences the collisions and their results. \citet{higa1996} analysed collisions of solid ice spheres with radius of $1.5\,\mathrm{cm}$ and decimetre ice blocks at different ambient temperatures ranging from $113\,\mathrm{K}$ to $269\,\mathrm{K}$. They found that the critical velocity between elastic and inelastic collisions remains constant at temperatures up to about $230\,\mathrm{K}$ and then decreases with increasing temperature. At $256\,\mathrm{K}$, the critical velocity is about a factor of 4 lower than for temperatures beneath $230\,\mathrm{K}$. \citet{higa1998} show that the fracture strength of solid ice has a very similar temperature dependence. Additionally, \citet{higa1996} show that the dependence of the coefficient of restitution on the impact velocity does not change with the ambient temperature, only the threshold to inelastic collision changes. Taking the results of these collision experiments into account, we assume that the collision results do not change at lower ambient temperatures, but the threshold conditions have to be scaled. This idea is supported by a comparison to the results of \citet{musiolik2015}, who analysed collisions of $\mathrm{CO_2}$ ice at temperatures of around $80\,\mathrm{K}$. With certain restrictions, the analytical model \citet{musiolik2015} develop for the coefficient of restitution fits quite well to our results (see Fig. \ref{rest}).\\
Collision velocities between centimetre and decimetre bodies in protoplanetary discs depend on the distance to the star and the disc model.  Close to the snowline \citet{brauer2008} calculate relative velocities in their numerical simulations of up to $45\,\mathrm{m\,s^{-1}}$. In their model, based on the MMSN, \citet{windmark2012a} calculate relative velocities of up to $20\,\mathrm{m\,s^{-1}}$. \citet{zsom2010} give an overview of different disc models and their influence on collision velocities. The velocities analysed in the experiments presented here fit very well to the range of velocities expected in the different disc models. The outcome of collisions are of importance especially for coagulation models, but also for models describing planetesimal formation by gravitational collapse. In the regions of high particle concentration, relative velocities are believed to be significantly lower. Therefore, the free fall collisions at velocities of up to $7\,\mathrm{m\,s^{-1}}$, especially the fragmentation threshold, are particularly interesting here. \citet{higa1998} find that the fracture strength of ice is a factor of 2 higher at low temperatures. We thus expect the threshold velocity to fragmentation of solid ice close to the snowline to be a factor of 2 higher than at the ambient temperatures in our experiments.\\
Mass transfer from a smaller to a larger body has previously only been observed in collisions of porous centimetre and decimetre dust agglomerates \citep{deckers2014}. Previous experiments on ice collisions at similar velocities have either taken smaller targets \citep{arakawa1995} or have been conducted using porous bodies \citep{shimaki2012a}. The experiments presented here show the possibility of a decimetre body growing in collisions with small projectiles, at velocities of up to $45\,\mathrm{m\,s^{-1}}$. However, in the multiple collisions on to the same target, we find that erosion starts at only a hand full of collisions. Yet, growth of the larger body is still possible, as collisions with the fragments produced in large amounts by the disruption of the projectile can in turn also lead to mass transfer to the target at velocities of up to $50\,\mathrm{m\,s^{-1}}$. The lower ambient temperature at the snowline could potentially mean mass transfer at even higher velocities, as the threshold to collisions without mass gain most probably also increases at lower temperatures. Moreover, the threshold to erosion might shift to higher specific energies, as the fracture strength increases at lower temperatures. On the other hand mass transfer in collisions of small irregular particles will most probably start at higher collision velocities. As there are no previous experiments with mass gain in collisions of water ice, it is not clear how lower ambient temperatures may affect the accretion efficiency.\\

\section{Conclusions}

In our experiments, we investigated the collision dynamics of decimetre bodies made up of solid water ice and find mass transfer from the centimetre projectile to the decimetre target up to an upper threshold velocity that depends on the size of the projectile. The projectiles get disrupted in the collisions producing a large number of fragments. Collisions of these fragments can also lead to mass transfer to the target, if the collision velocity is high enough. All things considered, we show that some decimetre icy bodies can grow in collisions with centimetre projectiles and the consecutive collisions with the projectile fragments at velocities of up to $50\,\mathrm{m\,s^{-1}}$, while others are eroded in a series of impacts.\\

\acknowledgments

We would like to thank the Deutsche Forschungsgemeinschaft (DFG) for their funding within the frame of the SPP 1385 "The first 10 Million Years of the Solar System -- A Planetary Materials Approach".\\

\bibliography{literature}

\end{document}